\documentclass[12pt]{article}
\usepackage{a4wide}
\usepackage{latexsym}

\usepackage{pslatex}
\usepackage[latin1]{inputenc}
\usepackage[T1]{fontenc}

\def\bq{\begin{eqnarray}}
\def\eq{\end{eqnarray}}

\def\v{\verb}

\newlength{\dinwidth} \newlength{\dinmargin}
\begin{document}
\thispagestyle{empty}

\begin{flushright}
  UPRF-2002-03
\end{flushright}

\vspace{1.5cm}

\begin{center}
  {\Large\bf Fast evolution of parton distributions\\}
  \vspace{1cm}
  {\large Stefan Weinzierl}\\
  \vspace{1cm}
  {\small {\em Dipartimento di Fisica, Universit\`a di Parma,\\
       INFN Gruppo Collegato di Parma, 43100 Parma, Italy}} \\
\end{center}

\vspace{2cm}

% abstract ---------------------------------------
\begin{abstract}\noindent
  {%
    I report on a numerical program for the evolution of parton distributions.
    The program uses the Mellin-transform method with an optimized
    contour. Due to this optimized contour the program needs only a few
    evaluations of the integrand and is therefore extremely fast.
    In addition, the program can also be used to reproduce the results
    of the x-space method.
   }
\end{abstract}

\vspace*{\fill}

\newpage 

{\bf\large PROGRAM SUMMARY}
\vspace{4mm}
\begin{sloppypar}
\noindent   {\em Title of program\/}: partonevolution \\[2mm]
   {\em Version\/}: 1.0 \\[2mm]
   {\em Catalogue number\/}: \\[2mm]
   {\em Program obtained from\/}: {\tt http://www.fis.unipr.it/\~{}stefanw/partonevolution} \\[2mm]
   {\em E-mail\/}: {\tt stefanw@fis.unipr.it} \\[2mm]
   {\em License\/}: GNU Public License \\[2mm]
   {\em Computers\/}: all \\[2mm]
   {\em Operating system\/}: all \\[2mm]
   {\em Program language\/}: {\tt C++     } \\[2mm]
   {\em Memory required to execute\/}: 
         negligible (1 MB)  \\[2mm]
   {\em Other programs called\/}: none \\[2mm]
   {\em External files needed\/}: none \\[2mm]
   {\em Nature of the physical problem\/}: 
         Parameterizations of parton distributions are usually given at a fixed
         input scale $Q_0$.
         To be used in perturbative calculations, they have to be evolved numerically
         to the desired scale $Q$.\\[2mm]
   {\em Method of solution\/}: 
         Inverse Mellin transform method.\\[2mm] 
   {\em Restrictions on complexity of the problem\/}: 
         None. \\[2mm]
   {\em Typical running time\/}:
         $10^{-2} s$ on a standard PC for an accuracy of $0.02 \%$, see also sect. \ref{sec:checks}.
\end{sloppypar}

% main text ------------------------------------
\newpage

\reversemarginpar

{\bf\large LONG WRITE-UP}

\section{Introduction}
\label{sec:intro}

Experiments at hadron colliders require apart from the hard scattering
amplitudes, which can be calculated in perturbation theory, also the
knowledge of the parton distributions functions (pdf's). Although these parton
distribution functions parameterize the non-perturbative information,
they depend on a factorization scale and the dependence on that scale
can be calculated within perturbation theory.
Usually parameterizations of parton distributions functions are given at
a low input scale \cite{Gluck:1998xa} - \cite{Martin:2001es} 
and then evolved to the desired scale.
At present the evolution kernel is known completely 
to next-to-leading order (NLO) \cite{Floratos:1979ny} - \cite{Hamberg:1992qt}.
The evolution has to be done numerically. There are two methods 
available: The first one consists in integrating numerically the
evolution equations in $x$ and in $Q^2$ and is called the x-space method.
The second one first performs a Mellin-transform on the 
variable $x$ \cite{Gluck:1976iz,Gluck:1990ze}.
The evolution equations then factorize and can be solved analytically.
Using this Mellin-transform method (or N-space method), leaves us with the
task to evaluate the inverse Mellin transform numerically. In this paper I
describe a numerical program, which uses an optimized integration contour
in the complex N-plane in order to do the inverse Mellin transform.
This method has been proposed by Kosower \cite{Kosower:1997hg}.
Using this method gives a fast and accurate program for the evolution
of parton distributions.
Although a variety of numerical evolution programs already 
exist (mainly within the x-space method) 
\cite{Miyama:1996bd} - \cite{Giele:2002hx},
they usually cannot compete in computation speed.
The required CPU time becomes an issue if one tries to extend
the analysis of experiments at hadron colliders in two important directions:
\begin{itemize}
\item Parton distributions with error-bars \cite{Giele:1998gw,Giele:2001mr}. 
The Tevatron data on the measurement
of the one jet inclusive transverse energy has shown \cite{Abe:1996wy}
that the method of obtaining
parton distributions from ``global fits'' has reached its limits.
Giele, Keller and Kosower \cite{Giele:1998gw,Giele:2001mr}
have proposed a method to include uncertainties into
the parton distributions.
This method involves a functional integration over a set of parton distributions.
With a set of pdf's containing 1000 or more parton distributions, the traditional
approach of a pre-calculated grid containing the evolved parton distributions
becomes inappropriate.
Instead, one would need a program which allows to calculate 
the evolution on the fly.
The program presented here allows one to do just that.
\item Extension to NNLO. The calculation of the three-loop anomalous dimensions
is currently under way \cite{Moch:2001fr,Moch:1999eb}. 
It is likely that the result will be rather lengthy.
The computational cost of evaluating the anomalous dimensions will therefore
be rather high, and one would like to minimize the number of function calls
to the anomalous dimensions.
The optimized contour employed in this program minimizes this number.
Once the three-loop anomalous dimensions are known, the program can be extended
in a straightforward way to include NNLO corrections.
\end{itemize}
The program is optimized for the N-space method, where one needs typically
only four or five evaluations of the anomalous dimensions to obtain the
evolution of 
a parton distribution with an accuracy of $2\cdot10^{-4}$.
In addition, I also include an option to simulate the x-space evolution programs.
x-space programs use a different truncation prescription and the results might
differ numerically from the N-space approach. It should be noted, that both
approaches are valid to next-to-leading order, the difference is formally
of higher order.
The x-space option is included to allow an easy comparison with those
programs, however it is not optimized.
In technical terms, since no closed formula for the evolution operator in the
singlet sector within the x-space approach is known, the evolution operator
is approximated by a sequence of evolutions over a small interval.
This slows down the program considerably.
The N-space evolution method is therefore the recommended one, where
the full performance of the program is obtained.\\
\\
This paper is organized as follows. 
In section \ref{sec:back} I recall the basic formulae
for the evolution of parton densities and sketch how to choose an optimized
contour using the Mellin transform method.
Section \ref{sec:design} gives an overview of the design of the program, while
section \ref{sec:howto} is of a more practical nature and describes how to install
and use the program.
In section \ref{sec:checks} I compare the program with the reference results
of Bl\"umlein et al. \cite{Blumlein:1996gv,Blumlein:1996rp} 
and address issues like performance.
A summary is provided in section \ref{sec:conclusions}.
The appendix collects the formulae for the evolution operators.

\section{Theoretical Background}
\label{sec:back}

The evolution equations for the parton distributions $f(x,Q^2)$ of the proton
read
\bq
\label{eq_1}
Q^2 \frac{\partial f(x,Q^2)}{\partial Q^2} & = & P(x,Q^2) \otimes f(x,Q^2),
\eq
where $x$ stands for the nucleon's momentum fraction carried by the parton,
$P(x,Q^2)$ is the Altarelli-Parisi evolution kernel, and $\otimes$
denotes the convolution
\bq
A(x) \otimes B(x) & = & \int\limits_0^1 dy \int\limits_0^1 dz
\delta(x- y z) A(y) B(z).
\eq
Eq. (\ref{eq_1}) is understood to represent the quark non-singlet evolution as well
as the coupled quark singlet and gluon evolution.
In the latter case $f$ is a two-component vector and $P(x,Q^2)$ a two-by-two
matrix.
Eq. (\ref{eq_1}) can be factorized by taking Mellin moments
\bq
A^z & = & \int\limits_0^1 dx \; x^{z-1} A(x)
\eq
and one obtains
\bq
\label{master_eq_Q2}
Q^2 \frac{\partial f^z(Q^2)}{\partial Q^2} & = & P^z(Q^2) \cdot f^z(Q^2).
\eq
Changing variables by using $a_s= \alpha_s(Q^2)/4\pi$ instead of $\ln Q^2$ as evolution variable
one obtains
\bq
\label{master_eq}
\frac{\partial f^z(Q^2)}{\partial a_s} & = & \frac{P^z(a_s)}{\beta(a_s)} f^z(Q^2),
\eq
where 
\bq
\beta(a_s) & = & Q^2 \frac{\partial a_s}{\partial Q^2}
\eq
is the beta function for the strong coupling.
Up to now, the evolution kernel is known completely only to next-to-leading order:
\bq
\label{trunc_P}
P^z(Q^2) & = & a_s P_0^z + a_s^2 P_1^z + O(a_s^3). 
\eq
To this order, the beta function is given by
\bq
\label{trunc_beta}
\beta(a_s) & = & - \beta_0 a_s^2 - \beta_1 a_s^3 + O(a_s^4).
\eq
Inserting eq. (\ref{trunc_P}) and eq. (\ref{trunc_beta}) into eq. (\ref{master_eq})
and expanding the r.h.s in $a_s$ one obtains
\bq
\label{master_eq_N}
\frac{\partial f^z(Q^2)}{\partial a_s} & = &
 - \frac{1}{\beta_0 a_s} 
    \left[ P_0^z + a_s \left( P_1^z - \frac{\beta_1}{\beta_0} P_0^z \right) \right]
    f^z(Q^2).
\eq
Solutions of eq. (\ref{master_eq_N}) are referred to as ``N-space'' solutions,
since this is the standard approach within the N-space method.
One the other hand, one might defer from inverting the power series in the denominator
on the r.h.s of eq. (\ref{master_eq}) and to solve eq. (\ref{master_eq}) directly.
To NLO, this amounts to solving
\bq
\label{master_eq_x}
\frac{\partial f^z(Q^2)}{\partial a_s} & = &
 - \frac{\left(P_0^z + a_s P_1^z \right)}{a_s \left( \beta_0 + \beta_1 a_s \right)} 
    f^z(Q).
\eq
Solutions of eq. (\ref{master_eq_x}) are referred to as ``x-space'' solutions,
again since this corresponds to the usual procedure within the x-space method.
Eq. (\ref{master_eq_N}) and eq. (\ref{master_eq_x}) differ by terms which are formally
of higher order in $a_s$, and solutions to both equations are therefore considered
accurate to next-to-leading order, although they might differ numerically.\\
\\
In addition there are different common practices on how to calculate $a_s$ at the scale $Q^2$.
At NLO $a_s(Q^2)$ can be found from the solution of the equation
\bq
\label{a_s_exact}
\frac{1}{a_s} - \frac{\beta_1}{\beta_0} \ln \left( \beta_1 + \frac{\beta_0}{a_s} \right)
 - \beta_0 L & = & 0,
\eq
where $L=\ln\frac{Q^2}{\Lambda_{QCD}^2}$.
This equation has to be solved numerically, and solutions of eq. (\ref{a_s_exact})
are referred to as ``exact'' NLO solutions.
On the other hand it is common practice to view $a_s$ as a power series
in $1/L$ and to use
\bq
\label{a_s_approx}
a_s & = & \frac{1}{\beta_0 L} \left( 1 -\frac{\beta_1}{\beta_0^2} \frac{\ln L}{L} \right)
\eq
to calculate $a_s$ at the scale $Q^2$.
Solutions of eq. (\ref{a_s_approx}) are referred to as ``truncated in $1/L$'' solutions.
Again, the numerical values might differ for solutions of eq. (\ref{a_s_exact})
and eq. (\ref{a_s_approx}).
It should be mentioned that the integration constant $\Lambda_{QCD}$ appearing
in eq. (\ref{a_s_exact}) is not identical to the integration constant of the same name
in eq. (\ref{a_s_approx}), e.g. fixing $a_s$ at a scale like $m_Z^2$ and solving
eq. (\ref{a_s_exact}) and eq. (\ref{a_s_approx}) for $\Lambda_{QCD}$ will
yield two different numerical values. Each value has to be used in conjunction with
the equation from which it was obtained.\\
\\
Eq. (\ref{master_eq_N}) and eq. (\ref{master_eq_x})
are formally solved by the introduction of an evolution
operator $E^z$:
\bq
f^z(Q^2) & = & E^z(a_s(Q^2),a_s(Q_0^2)) f^z(Q_0^2) 
\eq
Furthermore, there is yet another truncation method used in the literature:
Within this method one does not perform the change of the evolution variable
from $Q^2$ to $a_s$, but one substitutes the truncated solution eq. (\ref{a_s_approx}) 
for $a_s$ into eq. (\ref{master_eq_Q2}).
This amounts to solving
\bq
\label{master_eq_x_approx}
Q^2 \frac{\partial f^z(Q^2)}{\partial Q^2} & = & 
 \left( \frac{1}{\beta_0 L } \left( 1 -\frac{\beta_1}{\beta_0^2} \frac{\ln L}{L} \right)
          P_0^z
       + \frac{1}{\beta_0^2 L^2 } \left( 1 -\frac{\beta_1}{\beta_0^2} \frac{\ln L}{L} \right)^2
          P_1^z 
 \right) f^z(Q^2).
\eq
I will refer to solutions of eq. (\ref{master_eq_x_approx}) as
``x-space and truncated in $1/L$'' solutions.
Again, this equation is formally solved by an evolution operator, which now
depends on $Q^2$ and $Q_0^2$ instead of $a_s(Q^2)$ and $a_s(Q_0^2)$:
\bq
f^z(Q^2) & = & \hat{E}^z(Q^2,Q_0^2) f^z(Q_0^2) 
\eq
Formulae for the evolution operators corresponding to the various truncation prescriptions
are collected in the appendix.
The evolved parton distributions in x-space are then obtained by an inverse
Mellin transformation:
\bq
f(x,Q^2) & = & \frac{1}{2\pi i} \int\limits_C dz \; x^{-z} f^z(Q^2).
\eq
Here, the contour $C$ runs to the right of all singularities of the integrand.\\
The parton distributions are usually parametrized at the input scale $Q_0^2$ in a
form like
\bq
\label{parameterization}
x f(x,Q_0^2) & = & \sum\limits_{i} A_i x^{\alpha_i} (1-x)^{\beta_i} 
\eq
with Mellin transform
\bq
f^z(Q_0^2) & = & \sum\limits_{i} A_i B(z+\alpha_i-1,1+\beta_i),
\eq
where $B(x,y)$ is Euler's beta function.
Dropping from now on the arguments $Q^2$ and $Q_0^2$ our task is to
evaluate the integral
\bq
\label{inverse_mellin_int}
I & = & \mbox{Re} \; \frac{1}{\pi i} \int\limits_{C_s} dz \; E^z F(z), \nonumber \\
& & F(z) = x^{-z} \; \sum\limits_i A_i B(z+\alpha_i-1,1+\beta_i),
\eq
where complex conjugation has been used to replace the contour $C$ by $C_s$,
starting at the real axis, right to the right-most pole and running upwards
to infinity.
In the singlet case $F(z)$ and $A_i$ are vectors and $E^z$ is a matrix.
The most efficient way to solve this problem is to choose a contour in such a
way that the integrand can very well be approximated by some set of
orthogonal polynomials.
The method is due to Kosower \cite{Kosower:1997hg}.
It uses a parabolic contour, completely determined by the parton
distribution at the original scale $Q_0^2$ and evaluates the integral
by a Gauss-Laguerre quadrature formula.
With this method a few (e.g. four or five) evaluations of the integrand
are sufficient to evaluate the integral to very high precision 
(e.g. to an accuracy of 0.02 \%).\\
\\
I shortly review this method for the non-singlet case.
The contour of integration is given by
\bq
z(u) = z_0 + i c_2 \sqrt{u} + \frac{1}{2} c_2^2 c_3 u, \;\;\; u=0...\infty,
\eq
where $z_0$ is the minima of the function $F(z)$ on the real axis right
to the rightmost pole of $F(z)$.
The parameters $c_2$ and $c_3$ are given by
\bq
\label{eq_c2}
c_2 = \sqrt{\frac{2 F(z_0)}{F''(z_0)}}, 
& &
c_3 = \frac{F'''(z_0)}{3 F''(z_0)}.
\eq
$F'(z)$, $F''(z)$ and $F'''(z)$ are the first three derivatives of $F(z)$
with respect to $z$.
For the parameterization eq. (\ref{parameterization})
they are given by
\bq
F' & = & \sum\limits_{i} G_i' F_i, \nonumber \\
F'' & = & \sum\limits_{i} \left( G_i'' + G_i'^2 \right) F_i, \nonumber \\
F''' & = & \sum\limits_{i} \left( G_i''' + 3 G_i'' G_i' + G_i'^3 \right) F_i,
\eq
where $F_i = x^{-z} A_i B(z+\alpha_i-1,1+\beta_i)$
and
\bq
G_i & = & \ln ( x^{-z} B(z+\alpha_i-1,1+\beta_i)), \nonumber \\
G_i' & = & - \ln x + \psi(z+\alpha_i-1) -\psi(z+\alpha_i+\beta_i), \nonumber  \\
G_i'' & = &  \psi'(z+\alpha_i-1) -\psi'(z+\alpha_i+\beta_i), \nonumber \\
G_i''' & = & \psi''(z+\alpha_i-1) -\psi''(z+\alpha_i+\beta_i).
\eq
Here, $\psi$, $\psi'$ and $\psi''$ are polygamma functions.
With this parameterization one obtains for the integral eq. (\ref{inverse_mellin_int}):
\bq
I & = & \frac{c_2}{2 \pi} \int\limits_{0}^{\infty}
  \frac{du}{\sqrt{u}} \; e^{-u} \; \mbox{Re}
  \left[ e^{u} \left(1-i c_2 c_3 \sqrt{u} \right)
       E^{z(u)} F\left(z(u)\right) \right].
\eq
The function $u^{-1/2} e^{-u}$ is the weight function of the Laguerre
polynomials $L_k^{-1/2}(x)$ and the integral
can be approximated by a Gauss-Laguerre quadrature formula:
\bq
I & \approx & \frac{c_2}{2 \pi} \sum\limits_{j=1}^{k}
  w_j \; \mbox{Re}
  \left[ e^{u_j} \left(1-i c_2 c_3 \sqrt{u_j} \right)
       E^{z(u_j)} F\left(z(u_j)\right) \right].
\eq
Here, $u_j$ are the zeros of $L_k^{-1/2}(x)$ 
and the weights $w_j$ are given by
\bq
w_j & = & \frac{ \Gamma \left( n+\frac{1}{2} \right) }{n! (n+1)^2}
          \frac{u_j}{\left( L_{k+1}^{-1/2}(u_j) \right)^2}.
\eq
In the singlet case one has a two-component vector given by the
quark singlet combination $\Sigma^z$ and the gluon distribution $G^z$.
There will be strong mixing already for moderate evolution in $Q^2$, since $\Sigma^z$
and $G^z$ are not eigenfunctions of the evolution kernel.
One might be tempted to work in a basis of eigenfunctions 
of the leading-order evolution kernel and to extend 
the method discussed above to these eigenfunctions,
eventually choosing two different contours for the two eigenfunctions.
However this approach faces two problems:
First of all, one of the two eigenfunctions usually does not have
a minimum on the real axis on the right of the rightmost pole.
Secondly, by using different contours for the two components one is forced 
to keep track of branch cut crossings, a task one would like to avoid.
It is therefore better to use a single contour for both components,
although it might not be the optimal one for both components.\\
In the singlet sector the contour is chosen as follows
(this differs from the original proposition of Kosower):
$z_0$ is chosen as the minimum of the sum of the $\Sigma^z$- and the $G^z$-component.
I then diagonalize the leading-order evolution kernel at $z_0$:
\bq
\label{diag_z0}
\left( \begin{array}{c} e_1^z \\ e_2^z \end{array} \right)
& = &
\left( \begin{array}{cc} -\gamma_{0,qg}(z_0) & \gamma_{0,qq}(z_0)-\lambda_-(z_0) \\
        \gamma_{0,gg}(z_0)-\lambda_+(z_0) & -\gamma_{0,gq}(z_0) \end{array} \right)
\left( \begin{array}{c} \Sigma^z \\ G^z \end{array} \right),
\eq
where the eigenvalues $\lambda_{\pm}$ are given by 
\bq
\lambda_{\pm}(z_0) & = & \frac{1}{2} \left[
     \gamma_{qq}^{(0)}(z_0) + \gamma_{gg}^{(0)}(z_0)
  \pm \sqrt{ \left( \gamma_{gg}^{(0)}(z_0)-\gamma_{qq}^{(0)}(z_0) \right)^2
   +4 \gamma_{qg}^{(0)}(z_0) \gamma_{gq}^{(0)}(z_0) } \right].
\eq
Note that any $z$-dependence in eq. (\ref{diag_z0}) 
is only due to $\Sigma^z$ or $G^z$, $z_0$ 
appearing in the anomalous dimensions is a fixed parameter.
Strictly speaking the eigenvectors of eq. (\ref{diag_z0}) are 
only eigenvectors at the point $z=z_0$
of the leading-order evolution kernel.
I then take $e_1^z$ to determine $c_2$ and $c_3$ and integrate both components with
this contour. 
To compensate for the fact that $z_0$ is not necessarily the minimum of
$e_1^z$, the formula eq. (\ref{eq_c2}) is replaced by
\bq
c_2 & = & \sqrt{\frac{2 F(z_0) F''(z_0)}{F''(z_0)^2 - F'(z_0) F'''(z_0)}},
\eq
which takes into account corrections due to the fact, that the contour
does not start from a minimum of $F(z)$ (e.g. $F'(z_0) \neq 0$).
The reasons why this works is that in the cases of interest $e_1^z$ is 
roughly the sum of $\Sigma^z$ and $G^z$ and so $z_0$ is approximately optimal for
$e_1^z$. The other component $e_2^z$ is roughly the difference and numerically therefore
smaller.

\section{Design of the program}
\label{sec:design}

The program is written in C++.
All classes and functions are defined in the namespace \v/pdf/.
The parameterization of a parton distribution at the initial
scale $Q_0^2$ is represented by the class \v/partondistribution/.
A \v/partondistribution/ is constructed as follows:
\begin{verbatim}
  // u-valence CTEQ 4M
  int n = 2;
  double Q0 = 1.6;
  double A_u[2]     = { 1.344, 1.344*6.402};
  double alpha_u[2] = { 0.501, 0.501+0.873 };
  double beta_u[2]  = { 3.689, 3.689 };
  int eta = -1;

  partondistribution uvalence =
    partondistribution(n,Q0,A_u,alpha_u,beta_u,eta);
\end{verbatim}
This corresponds to the parameterization 
\bq
x \; u_v(x,Q_0^2) & =& 1.344 \; x^{0.501} (1-x)^{3.689} \left( 1 + 6.402\; x^{0.873} \right)
\eq
at the initial scale $Q_0 = 1.6 \; \mbox{GeV}$. Valence like distributions
need to have $\eta=-1$ explicitly in the constructor. For distributions which
correspond to $\eta=1$
(e.g. distributions in the singlet sector or non-singlet non-valence distributions
like $(u+\bar{u})-(d+\bar{d})$), the $\eta$-parameter can be dropped.\\
\\
This partondistribution can be evolved to a scale $Q$ by calling the function
\v/evolve_nonsinglet/:
\begin{verbatim}
  // evolution to 10 GeV for x = 0.01
  double Q = 10.0;
  double x = 0.01;

  double u_v = evolve_nonsinglet(uvalence,Q,x);
\end{verbatim}
The function \v/evolve_nonsinglet/ is just an entry point and the job is passed 
to the function
\v/evolve_nonsinglet_5/, which performs a Gauss-Laguerre quadrature with 5 points.
In addition there are functions \v/evolve_nonsinglet_3/, \v/evolve_nonsinglet_10/,\\
\v/evolve_nonsinglet_20/ and \v/evolve_nonsinglet_30/, corresponding to 
quadrature formulae with
3, 10, 20 and 30 points.
These functions can be selected by passing an additional argument to
\v/evolve_nonsinglet/, e.g.
\begin{verbatim}
  // evolution using a 10-point quadrature formula
  double u_v = evolve_nonsinglet(uvalence,Q,x,10);
\end{verbatim}
will call the function \v/evolve_nonsinglet_10/.\\
In the singlet sector evolution is done with the help of the function
\v/evolve_singlet/:
\begin{verbatim}
  // singlet evolution
  vector_d singlet = evolve_singlet(Sigma,Gluon,Q,x);

  double s = singlet[0];
  double g = singlet[1];
\end{verbatim}
\v/Sigma/ and \v/Gluon/ are of type \v/partondistribution/ and represent the
parameterization of the quark singlet combination and the gluon distribution, respectively.
The evolution function returns a two-component vector, with the evolved quark singlet
distribution as first component and the evolved gluon distribution as second component.
In C and C++ subscription of arrays starts at zero.
As in the non-singlet case the function \v/evolve_singlet/ is just an entry point, calling
\v/evolve_singlet_5/. Again, there are functions using quadrature formulae with 3, 5, 10,
20 and 30 points.\\
\\
In addition there are the classes
\v/evolutionkernel/ and \v/alpha_strong/.
The class\\
\v/evolutionkernel/ has a static data member \v/method/, which allows one
to select the truncation method, e.g.
\begin{verbatim}
  // select truncation method for evolution kernel
  evolutionkernel::method = evolutionkernel::N_space; 
\end{verbatim}
will perform the evolution according to eq. (\ref{master_eq_N}).
Other choices are \v/evolutionkernel::x_space/ and 
\v/evolutionkernel::x_space_truncate_in_one_over_L/,
corresponding to evolution according to eq. (\ref{master_eq_x}) and
eq. (\ref{master_eq_x_approx}), respectively.\\
\\
In a similar way, the class \v/alpha_strong/ has a static data member \v/method/,
which allows one to select the truncation method for $\alpha_s$, e.g.
\begin{verbatim}
  // select truncation method for alpha_strong
  alpha_strong::method = alpha_strong::truncate_in_one_over_L; 
\end{verbatim}
will use formula eq.(\ref{a_s_approx}), whereas the choice
\v/alpha_strong::exact/ will use eq. (\ref{a_s_exact}).
In addition the user should set the flavor thresholds and the values
of $\Lambda_{QCD}$ for the different number of flavors, as for example in
\begin{verbatim}
  // define values for alpha_strong (in GeV)
  alpha_strong::charm_threshold  = 1.3; 
  alpha_strong::bottom_threshold = 4.7; 

  alpha_strong::lambda_3 = 0.374; 
  alpha_strong::lambda_4 = 0.327; 
  alpha_strong::lambda_5 = 0.226; 
\end{verbatim}
Other classes of general interest provided by the package are the class
\v/complex_d/ corresponding to complex numbers with double precision
as well as the templates
\v/vector_template/
\v/<class T>/ 
and \v/matrix_template<class T>/.
The specializations \v/vector_d/, \v/vector_c/,\\
\v/matrix_d/ and \v/matrix_c/ define
vectors and matrices with real or complex entries in double precision.

\section{How to Use the Library}
\label{sec:howto}

In this section I give indications how to install and use the program library.
Compilation of the package will build a (shared) library called
\v/libpartonevolution/. The user will then link his own
programs against the library.

\subsection{Installation}

The program library can be obtained from\\
\\
{\tt \hspace*{12pt} http://www.fis.unipr.it/\~{}stefanw/partonevolution}\\
\\
After unpacking, the library for the evolution of parton distributions
is build by issuing
the commands
\begin{verbatim}
  ./configure 
  make
  make install
\end{verbatim}
There are various options which can be passed to the configure script,
an overview can be obtained with {\tt ./configure -{}-help}.\\
\\
After installation, the shell script \v/partonevolution-config/ can be used
to determine the compiler and linker command line options required
to compile and link a program with the partonevolution library.
For example, \v/partonevolution-config --cppflags/ will give the path to the header files
of the library, whereas \v/partonevolution-config --libs/ prints out the flags
necesarry to link a program against the library.

\subsection{Writing programs using the library}

Once the library is build and installed, it is ready to be used. Here 
is a small example program, which defines a parameterization at the scale
$Q_0 = 2 \;\mbox{GeV}$ and calls the evolution routine for $Q=10\;\mbox{GeV}$ and
$x=0.01$ .
\begin{verbatim}
#include <iostream>
#include "partonevolution/partonevolution.h"

int main()
{
  using namespace pdf;

  // select truncation methods
  alpha_strong::method = alpha_strong::truncate_in_one_over_L;
  evolutionkernel::method = evolutionkernel::N_space;

  // define a toy model with 4 flavors
  alpha_strong::charm_threshold  = 0.0;
  alpha_strong::bottom_threshold = 1E10;
  alpha_strong::lambda_4 = 0.250;

  // define a parton distribution
  double Q0 = 2.0;
  double A_u[1]     = { 35/16.};
  double alpha_u[1] = { 0.5 };
  double beta_u[1]  = { 3.0 };
  int eta = -1;

  partondistribution f =
    partondistribution(1,Q0,A_u,alpha_u,beta_u,eta);

  // evolution
  double Q = 10.0;
  double x = 0.01;

  double res = evolve_nonsinglet(f,Q,x);

  // print out result
  cout << "x*f(x,Q^2) = " << x*res << endl;
}
\end{verbatim}
After compilation and linking against the partonevolution library,
one obtains an executable, which will print out
\begin{verbatim}
x*f(x,Q^2) = 0.247242
\end{verbatim}
This program defines the $u_v$-distribution 
\bq
x u_v(x,Q_0^2) & = & \frac{35}{16} x^{0.5} (1-x)^{3.0}
\eq
of the toy model of ref. \cite{Blumlein:1996rp}.
The numerical difference of our value with 
the value 0.24723 quoted in ref. \cite{Blumlein:1996rp}
is well below the estimated accuracy given in ref. \cite{Blumlein:1996rp}.

\subsection{Documentation}

The complete documentation of the program is inserted as comment lines in
the source code.
The documentation can be extracted from the sources
with the help of the documentation system ``doxygen'' \cite{doxygen}.
The program ``doxygen'' is freely available.
Issuing in the top-level build directory for the partonevolution library the commands
\begin{verbatim}
  doxygen Doxyfile
\end{verbatim}
will create a directory ``reference'' with the documentation in html and latex format.

\section{Checks and performance}
\label{sec:checks}

In this section I compare the results of our program with the reference results 
of Bl\"umlein et al. \cite{Blumlein:1996rp}.
The authors of ref. \cite{Blumlein:1996rp} define a
toy model, consisting of four active
flavors with $\Lambda^{(4)}_{QCD}=250\mbox{MeV}$. At the input
scale $Q_0^2 = 4 \mbox{GeV}^2$ they take the initial parton distributions as 
follows:
\bq
x u_v = A_u x^{0.5} (1-x)^3, & & x d_v = A_d x^{0.5} (1-x)^4, \nonumber \\
x S = A_S x^{-0.2} (1-x)^7, & & x g = A_g x^{-0.2} (1-x)^5, \nonumber \\
x c = 0, & & x \bar{c} = 0. 
\eq
The sea $S$ is taken to be $SU(3)_{flavor}$-symmetric and to carry
$15 \%$ of the nucleon's momentum at the input scale.
This, together with the usual flavor- and momentum sum rules, fixes
the constants $A_u$, $A_d$, $A_S$ and $A_g$.
The sea is related to the quark singlet distribution
\bq
\Sigma = \sum\limits_{q=u,d,...} \left( q + \bar{q} \right)
\eq
by
\bq
S & = & \Sigma - u_v - d_v.
\eq
They provide a table with numerical results for the evolved parton distributions 
at the scale 
$Q^2 = 100 \;\mbox{GeV}^2$.  
They have estimated the numerical accuracy of their results to be $0.02 \%$.
I have compared the program with the contents of table 1 of \cite{Blumlein:1996rp}.
Using a 5-point quadrature formula the results of the program  
agree with the numbers presented there within the indicated estimated
accuracy (0.02\%), with some exceptions for the charm distribution and the singlet case 
for small values of $x$.
Using a 10-point quadrature formula for those cases
will lead to results within the $0.02\%$ uncertainty
margin.
I have compared both the x-space solution 
(upper part of the table 1 in ref. \cite{Blumlein:1996rp}, 
corresponding to solutions of eq. (\ref{master_eq_x_approx}) ) 
as well as the N-space solution 
(lower part of table 1 in ref. \cite{Blumlein:1996rp},
corresponding to solutions of eq. (\ref{master_eq_N}) 
in combination with eq. (\ref{a_s_approx}) for the strong coupling).\\
\\
Finally, I would like to give some indication on the required CPU time.
The evolution of a non-singlet parton distribution with a 5-point Gauss-Laguerre
quadrature formula takes about $1.3 \cdot 10^{-3} s$ on a standard PC (Athlon, 1.6 GHz).
The singlet evolution requires the evaluation of four anomalous dimensions.
In addition there is some overhead for vector and matrix arithmetic.
The singlet evolution is therefore
more expensive in terms of CPU time.
The evolution of the singlet combination takes about $6 \cdot 10^{-3} s$, again with
a 5-point Gauss-Laguerre quadrature formula on the same PC.
These numbers apply to the N-space method.
Simulating the x-space evolution method,
the evolution of a non-singlet parton distribution with a 5-point Gauss-Laguerre
quadrature formula takes about $1.3 \cdot 10^{-3} s$.
In the singlet sector, the evolution kernel has to be approximated by evolutions
over small intervals. Subdividing the evolution interval
into $1000$ subintervals, the evolution of the singlet combination 
takes about $270 \cdot 10^{-3} s$, again with
a 5-point Gauss-Laguerre quadrature formula on the same PC.

\section{Summary}
\label{sec:conclusions}

In this paper I described the program library ``partonevolution''.
This library can be used to evolve parton distributions given by
parameterizations at a scale $Q_0$ to the desired scale $Q$.
The program uses the Mellin transform method with an optimized contour.
Due to this optimized contour the program needs only a few
evaluations of the anomalous dimensions and is therefore extremely fast.
The library is therefore suited to be used in analysis involving parton
distributions with errors, where a functional integration over a set of
pdf's is needed.
Furthermore, once the three-loop anomalous dimensions are known, the extension
of the algorithms to NNLO is straightforward.
In addition, the program can also be used to simulate x-space evolution programs.

\begin{appendix}

\section{Evolution operators}

In this appendix the formulae for the evolution operators are collected.
I start with the N-space method. The evolution operator, which solves
eq. (\ref{master_eq_N}) is given in the non-singlet case 
by \cite{Furmanski:1982cw,Gluck:1990ze}
\bq
E^z_{\eta}\left(a_s(Q^2),a_s(Q_0^2)\right) & = & \left( \frac{a_s(Q^2)}{a_s(Q_0^2)}\right)^{\frac{\gamma^z_0}
{2 \beta_0}}
\left[ 1 + \frac{a_s(Q^2)-a_s(Q_0^2)}{2 \beta_0}
\left( \gamma^z_1(\eta) - \frac{\beta_1}{\beta_0} \gamma^z_0 \right) \right].
\eq
Here, $\eta=-1$ corresponds to the combinations $q - \bar{q}$, while
$\eta=+1$ should be used for the non-singlet combinations like
$(u+\bar{u})-(d+\bar{d})$.
I have expressed the evolution operator in terms of the anomalous dimensions
$\gamma_0^z$ and $\gamma_1^z$, which are related to the coefficients of the evolution kernel
by
$ P_i^z = - \frac{1}{2} \gamma_i^z$.
The explicit expressions for the anomalous dimensions 
can be found in \cite{Floratos:1981hs}.
For the singlet case one finds \cite{Furmanski:1982cw,Gluck:1990ze}:
\bq
E^z\left(a_s(Q^2),a_s(Q_0^2)\right) & = & \left( \frac{a_s(Q^2)}{a_s(Q_0^2)}\right)^{\frac{\lambda^z_-}
{2 \beta_0}}
\left[ P^z_- + \frac{a_s(Q^2)-a_s(Q_0^2)}{2 \beta_0} P^z_- \gamma^z P^z_- \right. \nonumber \\
& & \left. - \left( a_s(Q_0^2) - a_s(Q^2) \left( \frac{a_s(Q^2)}{a_s(Q_0^2)}\right)^{\frac
{\lambda^z_+-\lambda^z_-}{2 \beta_0}} \right)
\frac{P^z_- \gamma^z P^z_+}{2\beta_0 + \lambda^z_+ - \lambda^z_-} \right] \nonumber \\
& & + \left( + \leftrightarrow - \right)
\eq
with
\bq
\gamma^z & = & \gamma^z_1 -\frac{\beta_1}{\beta_0} \gamma^z_0, \nonumber \\
\lambda^z_{\pm} & = & \frac{1}{2}
\left( \gamma^z_{0,qq} + \gamma^z_{0,gg} \pm
 \sqrt{ \left( \gamma^z_{0,gg} -\gamma^z_{0,qq} \right)^2 + 4 \gamma^z_{0,qg} \gamma^z_{0,gq}
} \right), \nonumber \\
P^z_{\pm} & = & \pm \frac{1}{\lambda^z_+ - \lambda^z_-}\left(
\gamma^z_0 - \lambda^z_{\mp} \right).
\eq
The evolution operator solving eq. (\ref{master_eq_x}) (e.g. ``x-space method'')
reads in the non-singlet sector \cite{Kosower:1997hg}
\bq
E^z_{\eta}\left(a_s(Q^2),a_s(Q_0^2)\right) & = & \left( \frac{a_s(Q^2)}{a_s(Q_0^2)}\right)^{\frac{\gamma^z_0}
{2 \beta_0}}
\left( \frac{\beta_0+\beta_1 a_s(Q^2)}{\beta_0 +\beta_1 a_s(Q_0^2)} \right)^{\frac{\gamma^z(\eta)}{2 \beta_1}}.
\eq
In the singlet case no closed form is known and the evolution operator
has to be obtained as a sequence of evolutions over a small interval:
\bq
E^z\left(a_s(Q^2),a_s(Q_0^2)\right) 
 & = & E^z\left(a^{(n)},a^{(n-1)}\right) 
       E^z\left(a^{(n-1)},a^{(n-2)}\right) ... \; E^z\left(a^{(1)},a^{(0)}\right),
\eq
where $a^{(j)} = a_s(Q_0^2) + \left(a_s(Q^2)-a_s(Q_0^2)\right) j/n$.
In a small interval the evolution operator for the singlet case can 
be approximated by \cite{Kosower:1997hg}
\bq
\lefteqn{
E^z\left(a_s(Q^2),a_s(Q_0^2)\right) } \nonumber \\
& = & \left( \frac{a_s(Q^2)}{a_s(Q_0^2)}\right)^{\frac{\lambda^z_-}{2 \beta_0}}
\left[ P^z_- + \frac{1}{2 \beta_1} \ln\left( \frac{\beta_0+\beta_1 a_s(Q^2)}{\beta_0 +\beta_1 a_s(Q_0^2)} \right)
               P^z_- \gamma^z P^z_- 
         +  \frac{P^z_- \gamma^z P^z_+}{2\beta_0 + \lambda^z_+ - \lambda^z_-} \right. \nonumber \\ 
& & \left.    \left( a_s(Q^2) \left( \frac{a_s(Q^2)}{a_s(Q_0^2)}\right)^{\frac{\lambda^z_+-\lambda^z_-}{2 \beta_0}}
                      \,_2F_1\left(1,1+\frac{\lambda^z_+-\lambda^z_-}{2\beta_0},2+\frac{\lambda^z_+-\lambda^z_-}{2\beta_0},
                               -\frac{\beta_1}{\beta_0} a_s(Q^2) \right) \right. \right. \nonumber \\
& & \left. \left.
                   - a_s(Q_0^2) 
                      \,_2F_1\left(1,1+\frac{\lambda^z_+-\lambda^z_-}{2\beta_0},2+\frac{\lambda^z_+-\lambda^z_-}{2\beta_0},
                               -\frac{\beta_1}{\beta_0} a_s(Q_0^2) \right) 
              \right)
\right] \nonumber \\
& & + \left( + \leftrightarrow - \right).
\eq
Finally, the evolution operator for eq. (\ref{master_eq_x_approx}) (e.g. ``x-space and truncated
in $1/L$'') reads for the non-singlet case \cite{Kosower:1997hg}
\bq
\hat{E}^z_{\eta}(Q^2,Q_0^2) & = & \left(\frac{L_0}{L}\right)^{\frac{\gamma_0^z}{2 \beta_0}}
\exp\left[ -\frac{1}{2 \beta_0^2} \left( \frac{1}{L_0} -\frac{1}{L} \right) \left( \gamma_1^z(\eta) - \frac{\beta_1}{\beta_0} \gamma_0^z \right)
+\frac{\beta_1}{2 \beta_0^3} \left( \frac{\ln L_0}{L_0} - \frac{\ln L}{L} \right) \gamma_0^z \right. \nonumber \\
& & +\frac{\beta_1}{4 \beta_0^4} \left( \frac{1+2 \ln L_0}{L_0^2} - \frac{1+2 \ln L}{L^2} \right) \gamma_1^z(\eta) \nonumber \\
& & \left. -\frac{\beta_1^2}{54 \beta_0^6}
 \left( \frac{2+6 \ln L_0 +9 \ln^2 L_0}{L_0^3} - \frac{2+6 \ln L +9 \ln^2 L}{L^3} \right) \gamma_1^z(\eta) \right],
\eq
where $L_i = \ln Q_i^2 / \Lambda^2$.
In the singlet case no closed form is known and the evolution operator is
obtained as a sequence of evolutions
\bq
\hat{E}^z\left(Q^2,Q_0^2\right) 
 & = & \hat{E}^z\left(Q^2_{(n)},Q^2_{(n-1)}\right) 
       \hat{E}^z\left(Q^2_{(n-1)},Q^2_{(n-2)}\right) ... \; \hat{E}^z\left(Q^2_{(1)},Q^2_{(0)}\right),
\eq
where $Q^2_{(j)} = Q_0^2 + \left(Q^2-Q_0^2\right) j/n$.
In a small interval the evolution operator for the singlet case is then 
approximated by \cite{Kosower:1997hg}
\bq
\hat{E}^z(Q^2,Q_0^2) & = & \left( \frac{L_0}{L} \right)^{\frac{\lambda_-^z}{2 \beta_0}}
\left\{ P_-^z + P_-^z \left[
 -\frac{1}{2 \beta_0^2} \left( \frac{1}{L_0} -\frac{1}{L} \right)  \gamma^z
+\frac{\beta_1}{2 \beta_0^3} \left( \frac{\ln L_0}{L_0} - \frac{\ln L}{L} \right) \gamma_0^z \right. \right. \nonumber \\
& & +\frac{\beta_1}{4 \beta_0^4} \left( \frac{1+2 \ln L_0}{L_0^2} - \frac{1+2 \ln L}{L^2} \right) \gamma_1^z \nonumber \\
& & \left. -\frac{\beta_1^2}{54 \beta_0^6}
 \left( \frac{2+6 \ln L_0 +9 \ln^2 L_0}{L_0^3} - \frac{2+6 \ln L +9 \ln^2 L}{L^3} \right) \gamma_1^z \right]
P_-^z \nonumber \\
& & + P_-^z \gamma_1^z P_+^z \left[
 -\frac{1}{\beta_0 \left(2\beta_0+\lambda_+^z-\lambda_-^z\right)}
    \left(\frac{1}{L_0} - \frac{1}{L} \left( \frac{L_0}{L}\right)^{\delta^z} \right) \right. \nonumber \\
& & +\frac{2\beta_1}{\beta_0^3 \left(4 \beta_0+\lambda_+^z-\lambda_-^z\right)}
    \left(\frac{\ln L_0}{L_0^2} - \frac{\ln L}{L^2} \left( \frac{L_0}{L}\right)^{\delta^z} \right) \nonumber \\
& &  +\frac{2\beta_1}{\beta_0^2 \left(4 \beta_0+\lambda_+^z-\lambda_-^z\right)}
    \left(\frac{1}{L_0^2} - \frac{1}{L^2} \left( \frac{L_0}{L}\right)^{\delta^z} \right) \nonumber \\
& &  -\frac{\beta_1^2}{ \beta_0^5 \left(6 \beta_0+\lambda_+^z-\lambda_-^z\right)}
    \left(\frac{\ln^2 L_0}{L_0^3} - \frac{\ln^2 L}{L^3} \left( \frac{L_0}{L}\right)^{\delta^z} \right) \nonumber \\
& &  -\frac{2\beta_1^2}{\beta_0^4 \left(6 \beta_0+\lambda_+^z-\lambda_-^z\right)}
    \left(\frac{\ln L_0}{L_0^3} - \frac{\ln L}{L^3} \left( \frac{L_0}{L}\right)^{\delta^z} \right) \nonumber \\
& & \left. \left. -\frac{2\beta_1^2}{\beta_0^3 \left(6 \beta_0+\lambda_+^z-\lambda_-^z\right)}
    \left(\frac{1}{L_0^3} - \frac{1}{L^3} \left( \frac{L_0}{L}\right)^{\delta^z} \right)
\right] \right\}
 + \left( + \leftrightarrow - \right),
\eq
where $\delta^z=(\lambda_+^z - \lambda_-^z)/2\beta_0$. 
Note that there is a typo in eq. (8.27) of ref. \cite{Kosower:1997hg}. The 
prefactor $r_L^{-\lambda_-^z / \beta_0}$ should read $r_L^{+\lambda_-^z / \beta_0}$.

\end{appendix}

% -------------------------------

\end{document}